\documentclass{webofc}
\usepackage[varg]{txfonts}   
\usepackage{subcaption}

\begin{document}
\title{Network Capabilities for the HL-LHC Era}
%
%

\author{\firstname{Marian} \lastname{Babik}\inst{2} \and 
        \firstname{Shawn} \lastname{McKee}\inst{1} for the \\
        \firstname{HEPiX} \lastname{Network Function Virtualization Working Group\thanks{https://zenodo.org/record/3565562}} 
}

\institute{ Physics Department, University of Michigan, Ann Arbor, MI, USA
\and
            European Organisation for Nuclear Research (CERN), Geneva, Switzerland
          }

\abstract{%
High Energy Physics (HEP) experiments rely on the networks as one of the critical parts of their infrastructure both within the participating laboratories and sites as well as globally to interconnect the sites, data centres and experiments instrumentation. Network virtualisation and programmable networks are two key enablers that facilitate agile, fast and more economical network infrastructures as well as service development, deployment and provisioning. Adoption of these technologies by HEP sites and experiments will allow them to design more scalable and robust networks while decreasing the overall cost and improving the effectiveness of the resource utilization.

The primary challenge we currently face is ensuring that WLCG and its constituent collaborations  will have the networking capabilities required to most effectively exploit LHC data for the lifetime of the LHC. In this paper we provide a high level summary of the HEPiX NFV Working Group report that explored some of the novel network capabilities that could potentially be deployment in time for HL-LHC.

}
\maketitle
\section{Introduction}
\enlargethispage*{4mm}
\label{intro}

Network virtualisation and programmable networks are nowadays quite common in the commercial clouds and telecommunication deployments and have also been deployed by some of the Research and Education (R\&E) network providers to manage Wide-Area Networks (WAN). However there are only few HEP sites pursuing new models and technologies to build up their networks and data centers and most of the existing efforts are currently focused on improvements within a single domain or organisation, usually motivated by the organisation-specific factors. Therefore, most of the existing work is usually site or domain-specific. In addition, there is a significant gap in our understanding of how these new technologies should be adopted, deployed and operated and how the inter-play between LAN and WAN will be organised in the future. While it’s still unclear which technologies will become mainstream, {\it it’s already clear that software (software-defined) and programmable networks will play a major role in the mid-term.}

With the aim to better understand the technologies and their use cases for HEP a Network Functions Virtualisation Working Group (NFV WG) was formed within the High Energy Physics Information Exchange (HEPiX)\cite{HEPiX}. The group produced a report\cite{hepix_nfv_wg_report} identifying the work already done, looking at the existing projects and their results as well as better understanding the various approaches and technologies and how they might support HEP use cases.

This paper focuses on brief high level overview of the existing approaches in \textit{network virtualisation} and \textit{programmable networks}. It explains how current paradigm shift in the computing and clouds is impacting networking and how this will fundamentally change the ways networks are designed in the data centers and sites. \textit{Cloud native networking} approaches involving new topologies, network disaggregation and virtualisation have been identified as primary drivers that will impact data centre networking, which will in turn impact how data centres will be inter-connected in the future. In the second part which is devoted to the \textit{programmable wide-area networks}, capacity sharing, network provisioning and software-defined approaches where key R\&D projects in the area are highlighted.  The paper concludes with a proposed areas of future work and potential next steps. 



\section{Cloud Native Data Centre Networks}
\enlargethispage*{4mm}
One of the main drivers for network evolution in the data centres is the changing nature of the applications. With the dawn of virtualisation, applications have started to morph at an accelerating pace and moved from mostly static deployments (on bare metal servers) through virtual machines to containers and more recently to clusters of containers sometimes referred to as microservices. This evolution causes a particular change for networking, the usual life-cycle of the application (develop, deploy, update, re-deploy) has decreased from hours to microseconds. Establishing a full scale cluster of hundreds of containers can be done in less than a second, and upgrades or re-deployments of such a cluster can be done on a rolling basis, which means that the entire cluster can be replaced with new containers, all over again in seconds. Going even further, all this can be performed from a central location that can control a set of federated clusters requiring very little or no effort on the end sites to perform most of these tasks. This increasingly dynamic environment could become a major challenge for the networking infrastructure, which will need to keep up with this rapid pace.

The rise of Linux and the economics of scale has led to the development and operations of clouds. HEP sites have been predominantly statically deployed with allocations usually served by batch systems, operating at job level granularity with high capacity storage hosting the datasets locally. This is changing as experiments are starting to deploy their job payloads in containers and services are moving to container-based deployments (such as Kubernetes \cite{Kubernetes1}\cite{Kubernetes2}). This creates an interesting environment where multiple technologies are starting to overlap and compete. Some of the NSF-funded projects such as SLATE \cite{gardner2017slate}, that investigate new infrastructures for sites, are entirely based on Kubernetes. Physical analysis running in containers with full dataset uploaded to the cloud has been demonstrated running on the Google Cloud Platform \cite{Barisits:2648962}. It can be expected that this evolution will continue and accelerate, potentially having a major impact on the networking at HEP labs and sites.

Network engineers are facing major challenges while trying to accommodate the new computing models in an environment where they often need to support not only cutting edge container technologies that are now popular, but at the same time legacy systems ( “bare metal” ), virtual machines and other equipment that needs to be connected to the network or even multiple networks (e.g. experimental equipment, technical networks) with custom designs and protocols. 

Fast paced application life-cycle is not the only challenge, virtualisation is progressing into areas that were previously tied to the hardware, such as GPUs or storage systems. With GPU and storage virtualisation, there is a need for lower latency and higher throughput within the data centre in order to enable more efficient allocation and use of resources. 

Network vendors have already recognised the cloud opportunity and have started to re-profile their revenue expectations from enterprises towards cloud providers. This will likely have an impact on what the vendors will start offering and how the network infrastructure and supporting software will evolve in the mid-term.

\begin{figure}[t]
\centering
\includegraphics[width=\textwidth]{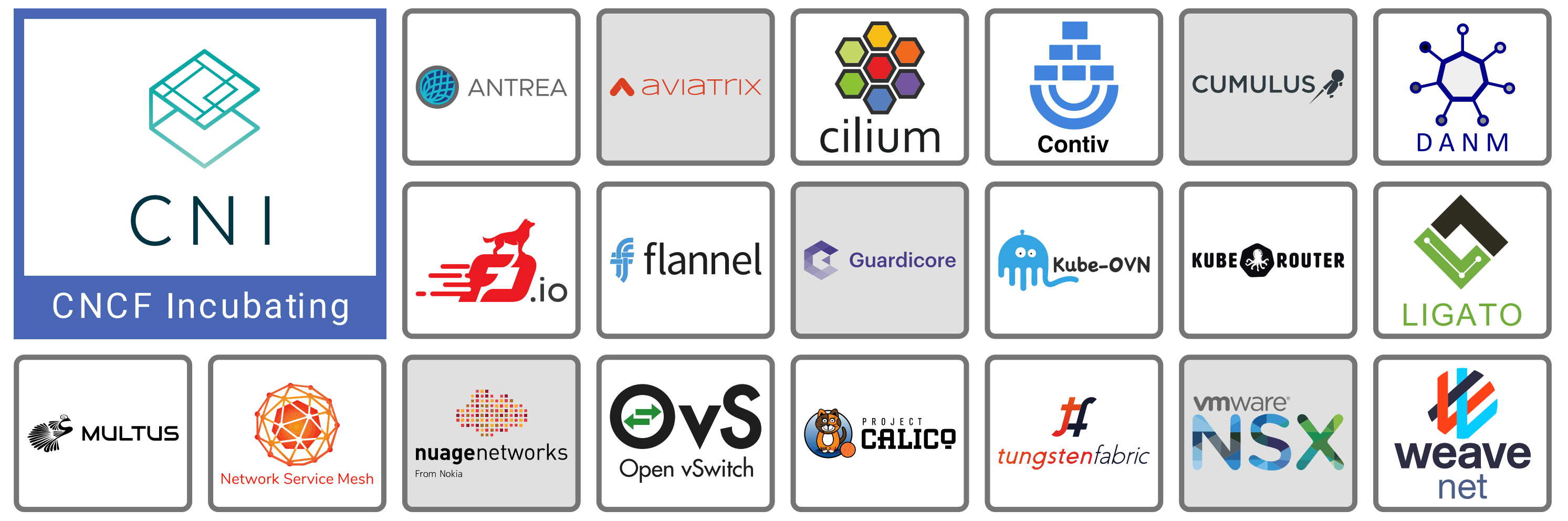}
\caption{CNCF Cloud Native Networking Landscape}
\label{fig-1}       
\vspace*{-15pt}
\end{figure}

\enlargethispage*{4mm}

The primary drivers that were identified in the report \cite{hepix_nfv_wg_report} as having a potentially strong impact on the data centre networks are following: 
\begin{itemize}
    \item \textbf{Container networking} - the current generation of applications are complex sets of services that run on a simple compute infrastructure with multiple levels of virtualisation that needs to rely on a simple networking model that scales easily and can support significant amount of east-west traffic. Deploying individual solutions for each functionality introduces complexity, making it extremely difficult to operate and troubleshoot. Coming up with a single solution is non trivial and requires both application and network engineers to come together, which (among other things) makes container networking hard. 
    \item \textbf{Rethinking network design} - current cloud-native data centers rely on the Clos topologies \cite{Clos} to build up a large-scale DC hosting container and VM technologies. Clos can be used to build very large networks with simple fixed form factor switches - allowing homogenous equipment - that greatly simplifies inventory management and configuration. The new interconnect model is usually based only on routing (layer-3) and bridging is supported only at the leaves (i.e. within a single rack). The rest of the inter-connectivity relies on some form of network virtualisation.
    \item \textbf{Network Virtualisation} - there are many existing approaches in network virtualisation, some of which are shown in Fig.\ref{fig-1}. They range from open source network operating systems running on hardware/bare metal switches and open routing platform, different software switch deployments up to Linux kernel network stack extensions. Currently it's unclear which approaches will become mainstream as a period of consolidation is likely coming. 
    \item \textbf{Network Disaggregation} - an important trend in network technologies that describes efforts to decouple network devices into open source hardware and open network operating systems. This will have a profound impact on the evolution of the network similar to how server disaggregation impacted compute in the past century. 
    \item \textbf{Programmable Network Interfaces} - an area of intense research and interest of network interface vendors, data center architects and end-users trying to optimize network performance. While network programmability in both NICs and switches has been possible in the past, significant advances were made recently due to network disaggregation efforts, invention of new programming languages and compilers (such as P4) as well as efficient hardware implementations. 
\end{itemize}

\enlargethispage*{4mm}
Cloud native networking and related technologies are providing a way how to design and cost effectively operate large scale DC networks. Still a number of challenges remain, both technological and non-technological that could impact a broader adoption by the HEP community:
\begin{itemize}
    \item Most existing HEP sites won’t be able to re-design their DC networking from scratch, requiring us to find ways to progressively migrate to new capabilities while accommodating existing constraints. 
    \item Historically, there has been a very clear separation between network and compute, but this no longer applies for the cloud native approaches where complex networking is present at the level of servers/hypervisors. This means that network and compute engineers must work together and build up expertise in the cross-domain areas. 
    \item Collaboration between the sites will also be very important to bridge the gap and come up with more effective approaches that better fit the existing HEP use cases. Encouraging closer collaboration between network and compute engineers within and across sites will be therefore an important factor in the adoption.
    \item Automation of the networking is another important area as relying on the open source network operating systems usually requires migrating to standard open source configuration tools. In addition, the usual approaches that work for configuration of compute might not work for network infrastructure due to various reasons. 
    \item Data Center Interconnect technologies (both HW and software-based) are quite novel approaches that will require initial deployments to evaluate how they could benefit inter-DC networking for federated use case such as data lakes \cite{WLCGdatalake}. 
    \item A range of other approaches that are only mentioned briefly in the report such as GPU, virtualized storage, hyper-converged architectures and edge services for HEP instrumentation and experiments will require initial testbeds and evaluations.
\end{itemize}

This section highlights core solutions and technologies that could help our community to rethink the way we design and operate our data centre networks and offer a great opportunity to build large scale data centers (centralised or distributed) that could benefit from economies of scale, simplification of the operational models and potential reduction in the overall operational cost, but apart from technology this will require new policies, priorities and funding to materialise. 

Some of the most promising areas of R\&D that could lead to a broader adoption of the mentioned technologies are container-based compute platforms such as SLATE edge services, HTCondor \cite{HTCondor} container back-ends or native container-based sites that could offer the best opportunity to test and evaluate cloud native networking. Provisioning of the storage servers with software switches and/or virtualized storage solutions is another area that has the potential to be easily deployed and tested. Finally, it’s also very important that cloud native approaches are considered for any planned extensions of existing centres or new data centers right from the start.

\section{Programmable Networks}
\enlargethispage*{4mm}
Paradigm shift in the computing and its impact on the network technologies as described in the previous section will make it easier to design and develop bigger data centres that will be inter-connected at very high capacities at lower latencies with a possibility to easily off-load to nearby Clouds, HPC centres or other opportunistic resources. A cluster of such centres can then create a federated site that will be exposed behind a single endpoint/interface for the experiments. This transformation has already started within the WLCG data lakes activities and some of the participating sites are already running their storage and or compute in a federated setup \cite{WLCGdatalake}. 

At the same time, small to medium sites will be able to complement the functionality of the core sites by off-loading some of their services by means of Kubernetes or other federated orchestrators. This can have a profound impact as design and development of the offline HEP distributed computing model can be radically simplified. HEP sites that could support different workloads by only running a single container orchestration or edge system are likely to be possible int he future (this is in a way revolutionary and will impact many different aspects of running a HEP site, apart from networking, also security, operations and management policies). 

From a network perspective operating a set of clustered DCs will bring its own challenges and will require closer collaboration with the R\&E providers. Provisioning of the networks, network telemetry, packet tracing and inspection as well as overall security and network automation will need to improve in order to make it easier for the federations to operate not only their inter-DC activities, but also easily expose their services to the outside.

Historically the National Research and Education Networks (NRENs) have managed to meet HEP networking needs by strategically purchasing capacity when network use exceeded trigger thresholds.  This has been a straightforward method to provide seemingly unlimited capacity for HEP, requiring no new technologies, policies or capabilities.  There were occasional “bumps” when regional or local capacities didn’t keep up, but overall over-provisioning resulted in an excellent networking for HEP. There are reasons to believe that the network situation will change due to both technological and non-technological reasons starting already in the next few years. Other data-intensive sciences will join with data scales similar to LHC\cite{Evans_2008}, which will impact not only R\&E providers, but also the way end-users are currently utilising the network. In the new multi-science high throughput environment, network provisioning, design and operations will need to evolve to better share and organise the available resources.

\enlargethispage*{4mm}
\textit{WAN programmable networks} address many of the challenges outlined above and have the potential to change the way HEP sites and experiments connect and interact with the network. Some of the key projects in the domain of orchestration, automation and virtualisation of WAN are following:
\begin{itemize}
    \item \textit{Programmable Networks for Data-Intensive Sciences} has a number of key technologies and projects in different areas such software-defined WAN (SD/WAN), software-defined exchanges (SDX), network orchestrators (e.g., SENSE\cite{SENSE} and NOTED\cite{NOTED}), network provisioning systems (e.g. multi-ONE) and network-aware data transfer systems (e.g. BigDataExpress\cite{BigDataExpress}. 
    \item Research \& Education Networks Programmable Services are planned by both ESNet\cite{ESnet} and GEANT\cite{GEANT} and include projects such as ESNet6\cite{ESnet6}, FABRIC\cite{FABRIC}, GEANT OAV\cite{GEANT-OAV} and GTS\cite{GEANT-GTS}. 
\end{itemize}

\subsection{Challenges and Outlook}
\enlargethispage*{4mm}
Programmable WAN is still an area of intensive research and development and while the existing projects have well defined scope and good match to the HEP use cases, there are still a number of challenges that remain: 
\begin{itemize}
\item One of the core challenges for some time is the fact that it appears to be difficult to bring the existing projects from testbed/prototype stage into production. Within LHCONE\cite{lhcone} R\&D efforts, a number of projects were successfully demo-ed in the past, but it has proven to be very challenging to deploy them in the production infrastructure. What appears to be missing are network infrastructures where prototypes can be tested at scale and then easily deployed/migrated to production (e.g. FABRIC).
\item Network provisioning will need to evolve to address multi science domain entering the R\&E networks requiring advances in capacity organisation, network management, accounting and monitoring.
\item There is currently a significant lack of available telemetry, tracing and insight into how the current network operate and how such data can be programmatically accessed. 
\item From the perspective of data transfer systems,  there are significant gaps in both achieving the maximum bandwidth (end-to-end) as well as organising allocations of capacity in ways that would avoid bottlenecks and allow more efficient sharing of the available capacity. 
\item As another alternative, automated methods for traffic engineering that would automatically adapt to the existing workloads have been proposed by different projects (both in SDX and orchestrators). Such systems promise to keep the existing status quo where state of the underlying network and its operations are transparent to the experiments. It remains to be seen if such approaches will be feasible in a large scale federated environments (such as LHCONE). 
\end{itemize}

\section{Proposed Areas of Future Work}
\enlargethispage*{4mm}
A primary goal of the HEPiX report \cite{hepix_nfv_wg_report} was to seed a collaboration between the experiments, the sites and the research and education networks to deliver capabilities needed by HEP for their future infrastructure while enabling the sites and NRENs to most effectively support HEP with the resources they have.
 
In this section we outline three possible areas of future work that can help tie together activities within and among the experiments and sites with network engineers, NRENs and researchers.  It is critical that we identify projects that are useful to the experiments, deployable by sites, and that involve a range of participants spanning the sites, the experiments and the (N)RENs.  Without the involvement of each, we risk creating something unusable, irrelevant or incompatible.
The following are not meant to be exclusive, merely suggestions based upon the working groups interactions and discussions amongst its members. 
\begin{itemize}
    \item \textbf{Making our network use visible}  - Understanding the HEP traffic flows in detail is critical for understanding how our complex systems are actually using the network.   
    With a standardized way of marking traffic, any NREN or end-site could quickly provide detailed visibility into HEP traffic to and from their site, a benefit for NRENs and users.  
    \item \textbf{Shaping data flows} - It remains a challenge for HEP storage endpoints to utilize the network efficiently and fully.  Shaping flows via packet pacing  to better match the end-to-end usable throughput results in smoother flows which are much friendlier to other users of the network by not bursting and causing buffer overflows. 
    \item \textbf{Network orchestration to enable multi-site infrastructures} - Within our data centers, technologies like OpenStack and Kubernetes are being leveraged to create very dynamic infrastructures to meet a range of needs. Critical for these technologies is a level of automation for the required networking using both software defined networking and network function virtualization.   
    As we look toward HL-LHC, the experiments are trying to find tools, technologies and improved workflows that may help bridge the anticipated gap between the resources we can afford and what will actually be required to extract new physics from massive data we expect to produce.  
    To support this type of resource organization evolution, we need to begin to prototype and understand what services and interactions are required from the network.  We would suggest a sequence of limited scope proof-of-principle activities in this area would be beneficial for all our stakeholders. 
\end{itemize}
\enlargethispage*{4mm}
\section{Conclusion and Summary}
We have described the work of the HEPiX Network Function Virtualization working group and their phase I report and indicated what we believe are the important areas for HEP to consider regarding future networking requirements as well as outlining specific proposed areas of work for the near, mid and long term.

\section{Acknowledgements}
\enlargethispage*{4mm}
We gratefully acknowledge the National Science Foundation which supported this work through NSF grants \#1836650 and \#1827116.   In addition, we acknowledge our collaborations with the CERN IT, WLCG and LHCONE/LHCOPN communities who also participated in this effort.

%
\bibliography{bibliography}

\begin{thebibliography}{20}

\bibitem{HEPiX}
HEPiX-Team, (2020), \emph{High-energy physics information exchange}, Retrieved
  from \urlstyle{tt}\url{https://www.hepix.org/ [accessed 2020-07-19]}

\bibitem{hepix_nfv_wg_report}
{HEPiX NFV Working Group}, Tech. rep. (2019),
  \urlstyle{tt}\url{https://doi.org/10.5281/zenodo.3565562}

\bibitem{Kubernetes1}
B.~Burns, B.~Grant, D.~Oppenheimer, E.~Brewer, J.~Wilkes, ACM Queue
  \textbf{14}, 70 (2016)

\bibitem{Kubernetes2}
\emph{Kubernetes | production-grade container orchestration} (2018),
  \urlstyle{tt}\url{https://kubernetes.io/}

\bibitem{gardner2017slate}
R.~Gardner, J.~Breen, L.~Bryant, S.~McKee, \emph{SLATE and the Mobility of
  Capability} (2017), \urlstyle{tt}\url{http://par.nsf.gov/biblio/10064986}

\bibitem{Barisits:2648962}
M.~Barisits, F.~Barreiro, T.~Beermann, K.~De, A.~Dubreuil, J.~Elmsheuser,
  A.~Klimentov, M.~Lassnig, P.~Love, T.~Maeno et~al. (ATLAS Collaboration),
  Tech. Rep. ATL-SOFT-PROC-2018-034, CERN, Geneva (2018),
  \urlstyle{tt}\url{https://cds.cern.ch/record/2648962}

\bibitem{Clos}
C.~{Clos}, \emph{A study of non-blocking switching networks} (1953), Vol.~32,
  pp. 406--424, ISSN 0005-8580

\bibitem{WLCGdatalake}
I.~Bird, S.~Campana, M.~Girone, X.~Espinal, G.~McCance, J.~Schovancová,
  \emph{Architecture and prototype of a WLCG data lake for HL-LHC} (2019), Vol.
  214, p. 04024

\bibitem{HTCondor}
D.~Thain, T.~Tannenbaum, M.~Livny, Concurrency and Computation: Practice and
  Experience \textbf{17}, 323 (2005),
  \texttt{https://onlinelibrary.wiley.com/doi/pdf/10.1002/cpe.938}

\bibitem{Evans_2008}
L.~Evans, P.~Bryant, \emph{{LHC} Machine} ({IOP} Publishing, 2008), Vol.~3, pp.
  S08001--S08001,
  \urlstyle{tt}\url{https://doi.org/10.1088%2F1748-0221%2F3%2F08%2Fs08001}

\bibitem{SENSE}
I.~Monga, (2020), \emph{{SDN} for end-to-end networking @ exascale}, Retrieved
  from
  \urlstyle{tt}\url{http://es.net/assets/pubs_presos/SENSE-Thomas-20160217-on-Web.pdf
  [accessed 2020-07-19]}

\bibitem{NOTED}
{Martelli, Edoardo}, {Manzi, Andrea}, {Keeble, Oliver}, {Cass, Tony},
  {Busse-Grawitz, Coralie}, To appear in proceedings of CHEP 2019, EPJ Web
  Conf.  (2020)

\bibitem{BigDataExpress}
Q.~{Lu}, L.~{Zhang}, S.~{Sasidharan}, W.~{Wu}, P.~{DeMar}, C.~{Guok},
  J.~{Macauley}, I.~{Monga}, S.~{Yu}, J.H. {Chen} et~al., \emph{BigData
  Express: Toward Schedulable, Predictable, and High-Performance Data
  Transfer}, in \emph{2018 IEEE/ACM Innovating the Network for Data-Intensive
  Science (INDIS)} (2018), pp. 75--84

\bibitem{ESnet}
ESnet-Team, (Feb 2020), \emph{Energy sciences network}, Retrieved from
  \urlstyle{tt}\url{http://www.es.net/ [accessed 2020-07-19]}

\bibitem{GEANT}
GEANT-Team, (Feb 2020), \emph{{GÉANT} is the leading collaboration on
  e-infrastructure and services for research and education.}, Retrieved from
  \urlstyle{tt}\url{http://geant.org/}

\bibitem{ESnet6}
J.~Metzger, (2020), \emph{{ESnet6}: an entirely new software-driven network
  design that enhances the ability to rapidly invent, test, and deploy new
  innovations.}, Wikipedia, Retrieved from
  \urlstyle{tt}\url{https://en.wikipedia.org/wiki/Energy_Sciences_Network#ESnet6}

\bibitem{FABRIC}
I.~Baldin, A.~Nikolich, J.~Griffioen, I.~Monga, K.C. Wang, T.~Lehman, P.~Ruth,
  \emph{FABRIC: A National-Scale Programmable Experimental Network
  Infrastructure} (2020), Vol.~23,
  \urlstyle{tt}\url{http://par.nsf.gov/biblio/10132161}

\bibitem{GEANT-OAV}
GEANT-Developers, (2019), \emph{D6.2 automation and orchestration of services
  in the {GEANT} community}, Web document, Retrieved from
  \urlstyle{tt}\url{https://www.geant.org/Projects/GEANT_Project_GN4-3/Pages/GN4-3_Deliverables.aspx\hspace
  [accessed 2020-07-19]}

\bibitem{GEANT-GTS}
GEANT-Developers, (2020), \emph{{GÉANT} testbeds service}, Web Page, Retrieved
  from
  \urlstyle{tt}\url{https://www.geant.org/Services/Connectivity_and_network/GTS/Pages/Home.aspxv[accessed
  2020-07-19]}

\bibitem{lhcone}
E.~Martelli, S.~Stancu, \emph{LHCOPN and LHCONE: Status and Future Evolution}
  (2015), Vol. 664, p. 052025,
  \urlstyle{tt}\url{http://stacks.iop.org/1742-6596/664/i=5/a=052025}

\end{thebibliography}
%
%

\end{document}